# Electrical Switching Dynamics in Circular and Rectangular $Ge_2Sb_2Te_5$ Nanopillar Phase Change Memory Devices

O. Ozatay, B. Stipe, J. A. Katine, and B. D. Terris
*Hitachi Global Storage Technologies-San Jose Research Center*
*San Jose CA 95135*

*ABSTRACT*

We have measured the critical phase change conditions induced by electrical pulses in $Ge_2Sb_2Te_5$ nanopillar phase change memory devices by constructing a comprehensive resistance map as a function of pulse parameters (width, amplitude and trailing edge). Our measurements reveal that the heating scheme and the details of the contact geometry play the dominant role in determining the final phase composition of the device such that a non-uniform heating scheme using rectangular contacts promotes partial amorphization/crystallization for a wide range of pulse parameters enabling multiple resistance levels for data storage applications. Furthermore we find that fluctuations in the snap-back voltage and set/reset resistances in repeated switching experiments are related to the details of the current distribution such that a uniform current injection geometry (i.e. circular contact) favors more reproducible switching parameters. This shows that possible geometrical defects in nanoscale phase change memory devices may play an essential role in the performance of the smallest possible devices through modification of the exact current distribution in the active chalcogenide layer. We present a three-dimensional finite element model of the electro-thermal physics to provide insights into the underlying physical mechanisms of the switching dynamics as well as to quantitatively account for the scaling behaviour of the switching currents in both circular and rectangular contact geometries. The calculated temporal evolution of the heat distribution within the pulse duration shows distinct features in rectangular contacts providing evidence for locally hot spots at the sharp corners of the current injection site due to current crowding effects leading to the observed behaviour.

**INTRODUCTION**

Following the original work of Ovshinsky[1], the reversible switching phenomena observed in disordered semiconductors has attracted substantial theoretical[2,3,4,5,6,7] and experimental[8,9,10,11,12,13] efforts targeted at gaining a deep understanding of the physical principles that govern the switching dynamics, and also learning how to control and optimize the power requirements coupled with characteristic switching timescales. Dramatic and ultra-fast (nanosecond scale) changes in the physical properties such as electrical resistivity and optical reflectivity upon amorphization or crystallization makes chalcogenides ideal potential candidates for a universal non-volatile memory device, especially if, the device switching characteristics are highly scalable to nanometer dimensions. The principle of phase-dependent optical reflectivity with large changes induced by local heating through laser pulses, has already been exploited in commercial optical storage media such as DVDs. On the other hand, the practical utilization of three to four orders of magnitude changes in the electrical resistivity induced by local joule heating through electrical pulses necessitates a detailed experimental characterization and theoretical modeling of the critical phase change conditions and their scaling properties.

In this work we use a resistance mapping technique whereby we apply electrical pulses of various amplitudes, widths and trailing edges and deduce the resulting phase composition in nanoscale $Ge_2Sb_2Te_5$ nanopillar devices. Such resistance map measurements, when repeated for devices with different contact sizes and geometries, provide a detailed picture of the scaling behaviour of the critical phase change conditions and the associated switching dynamics. A comparison of the observed characteristics with a three-dimensional finite element model (3D-FEM) of the electro-thermal physics enables a good assessment of the resulting device performance while providing physical insights into the nature of resistance switching behaviour.

## DEVICE STRUCTURE AND I-V CHARACTERISTICS

The device structure under study consists of a 50nm thick $Ge_2Sb_2Te_5$ (GST) chalcogenide phase change material sandwiched between two 50 nm $W_{0.8}Ti_{0.2}$ (WTi) heating element layers. The as-deposited GST film was annealed at ~300 °C to establish the polycrystalline phase throughout the film initially. A schematic drawing is shown in Fig.1 (a). The top WTi electrode is patterned into a nanopillar by using advanced e-beam lithography and ion-mill etching techniques that stop within a nm of the top GST-WTi interface utilizing a secondary-ion-mass-spectroscopy (SIMS) signal. This avoids ion-mill damage on the phase change material in an attempt to preserve bulk film properties. The electrical contacts to the device are made through Ta (5nm)/Au (~175nm) bilayers. All the devices are isolated by an alumina field-insulation layer. The active phase transition region is defined by the top GST-WTi interface where the local concentration of the applied current enables effective heating above the melting point whereas both the top and bottom metallic electrodes act as efficient heat sinks to quickly absorb the heat during cooling. Fig. 1 (b) shows a cross-sectional scanning electron microscope (SEM) image of a 75 nm diameter nanopillar. It is clearly observable that the actual contact area at the top GST-WTi interface can be 20-30% larger than the pillar diameter due to tapering as a result of ion-mill re-deposition effects, which defines a ~20-30% uncertainty in the contact area.

We have fabricated devices with linear dimensions (diameter or side) in the range of 50 nm to 1 μm and with rectangular and circular pillar contact geometries. Fig. 2 shows the current-voltage (IV) characteristics of both amorphous (in blue) and crystalline (in red) states of a 600 nm square device. The crystalline (SET) state exhibits a low resistance (~90Ω) accompanied by ohmic IV characteristics whereas the amorphous (RESET) state resistance at low bias (<~0.6V) is of the order of 10s of kΩs. Increasing the voltage bias above a threshold value (~0.65V in this case) a snap-back occurs as a result of formation of highly conductive paths in the amorphous material (also known as Ovonic threshold switching (OTS)). Since, after the snap-back, the material is still in the amorphous state this effect is believed to be of electronic origin (not the result of a structural change). The crystalline GST has acceptor like traps due to vacancies whereas the amorphous phase has both donor like and acceptor like traps due to dangling bonds in the Te sites. According to a model by Pirovano et al.[6] impact ionization leads to recombination in the shallow donor like traps pushing the electron quasifermi level closer to the conduction band. As the traps are saturated, the snap-back occurs and the conduction is dominated by impact ionization and electron transport. A different model by Ielmini et al.[14] suggests, electrons make it to the shallow traps either by thermal emmision or tunneling. The increasing conductivity causes band bending and the high current is sustained by tunneling through a thin off layer. While the physical origin of the snap-back is still not very well understood, from an application point of view it is the key process that enables phase-change memory operation by allowing the flow of electrical currents high enough to induce phase-change in the amorphous material. After the snap-back, substantial joule heating triggers crystal nucleation and growth causing an accelerated transformation to the thermodynamically favorable crystalline (SET) state via Ovonic memory switching (OMS).

The reset operation is performed by applying pulse amplitudes large enough for the GST to be locally heated above its melting point (~620 °C) and then super-cooling (melt-quenching) it below its glass transition temperature (~300 °C) into the amorphous phase. The switching back to the crystalline state can be achieved by sweeping the dc bias beyond the snap-back voltage as described above or alternatively by applying pulses of amplitude insufficient to melt but above the threshold for triggering crystal nucleation and growth. The former method requires the use of a set current compliance to protect the smallest devices, since excessive heating after OTS can result in device failure. Five such consecutive switching events following a reset pulse are shown in Fig. 3 (a) and (b) corresponding to a 90nm square and a 75 nm diameter circular contact respectively. The presence of parabolic (non-ohmic) IV characteristics in the set state (red curves) implies incomplete crystallization (i.e. residual amorphous content). Inhomogeneities in the phase composition can be visualized as an amorphous matrix in a crystalline background, which during the reset pulse, will only be influenced by heat diffusion from the surroundings, potentially resulting in temperatures not exceeding the melting point. The resulting fluctuations in the initial phase composition and resistance lead to variations in the snap-back voltage from one IV sweep to another. Figures 3 (a) and (b) suggest that this effect is more dominant for a square contact than for a circular contact, since the sharp corners in that case lead to current-crowding effects[15] and locally inhomogeneous distributions of temperature gradients (see the discussion below on finite-element calculations).

According to the percolation model of snap-back[16] there exists a percolation threshold at which the memory space is filled by high-mobility holes emitted from low-mobility charge traps due to the application of a threshold force of electric field such that a continuous cluster of conductive cells form between the top and bottom electrodes forming a percolation path. The measured values of the snap-back voltage (1.33 ± 0.07 V for the 90 nm square device and 1.13 ± 0.01 V for the 75 nm circular device as indicated in fig. 3) correspond to a threshold electric field of $2.66*10^5$ V/cm and $2.26*10^5$ V/cm for the square and circular devices respectively. The percolation threshold is given by[16]:

$$P_{th} = A.\exp(\frac{(\Lambda.F_{th})^{\gamma} - B}{kT}) \qquad (1)$$

where A is the scale coefficient in grained model, $\Lambda$ is the free path length, $F_{th}$ is the threshold electric field, $\gamma$ is a model parameter, B is the ionization barrier, k is the Boltzmann constant and T is the temperature. Using the measured threshold electric fields and the experimental values[17] of B = 0.3 ± 0.05 eV, $\Lambda$= 10 ± 4nm for GST, with A=1 and $\gamma$=1 yields a nominal percolation threshold of 0.3 ± 0.2 and 0.065 ± 0.005 for the square and circular samples respectively. This large difference in the percolation thresholds and their uncertainities indicates that a non-uniform heating scheme in rectangular contacts may affect the snap-back threshold by changing the percolation path configuration in each switching event where the percolation threshold is a sensitive indicator of initial phase-distribution per switching event. The large variation in the amorphous state resistance of the square device (fig 3 (a) inset) as opposed to marginal variation in the amorphous state resistance of the circular device (fig. 3 (b) inset) provides evidence for more prominent changes in the phase distribution and percolation paths in the square device. (As a side note, the measured set state resistance above the compliance threshold (at around 0.4V) is determined by the compliance current and therefore does not represent the actual device resistance. For this reason the dotted horizontal lines in the insets of fig. 3 (a) and (b) should be used as an aid to the eye for a rough estimate of the resistance window in this case). We have observed such strong variations in the snap-back voltage and amorphous state resistance for repeated switching events, only for rectangular devices in the sub-100nm contact size range. This could be a source of device failure in the smallest possible phase change memory cells where the device performance becomes very sensitive to any possible geometrical defects in the contact area.

**PULSED SWITCHING DYNAMICS**

Figure 4 displays repeated measurements of the pulsed reset operation to prepare a 75 nm circular device in the amorphous phase. The applied electrical pulse is characterized in terms of its width between 50% points, amplitude between the maximum and minimum points and the trailing (falling) edge as defined to be the time it takes for the voltage to drop from 90% to 10% of its amplitude as illustrated at the inset of fig. 4. The typical values of pulse widths and trailing edges used for successful reset operations are 50 to 100 ns and 2 ns respectively. The resulting value of resistance change is two to three orders of magnitude. The measured value of the threshold voltage for the reset operation for this 75 nm diameter circular device is approximately 0.6V. Since the typical snap-back threshold is above 1V (from fig. 3)( i.e larger than the reset threshold) we conclude that in this contact size scale the snap-back process immediately allows large enough currents (in the absence of a current limiter) to cause local melting in the GST layer. For this reason a set compliance is used to protect overheating of the device in the set operation by dc voltage sweeps. Similarly in pulsed switching experiments we used a 1kΩ series load resistor to avoid damage to the device due to excess currents. However, the lack of impedance matching between the load resistor and the output of the pulser (50Ω) leads to a 1/10 attenuation in the transmitted pulse amplitude with no detectable pulse distortion. Further multiple pulse reflections/distortions between the GST device and the 1kΩ surface-mount resistor were avoided by placing them in close proximity (within 3 cm). We note that using a load resistor as well as using a set compliance in combination with the parasitic capacitance of the GST device, can lead to unwanted programming transients and pulse distortions which can affect the device performance[18]. The measured lead capacitance in our samples was less than 2 pF providing a 2ns limit for the smallest achievable pulse trailing edge. All the pulse parameters reported here correspond to the actual values that appear on the GST device.

Repeating the reset switching experiments in fig. 4 as a function of pulse width and amplitude for a given trailing edge value we obtain phase diagrams as shown in fig. 5, which displays the results of a resistance map characterization of the pulsed reset switching of a 75 nm circular device in (a), (b) and (c) (with 2 ns, 22ns and 42 ns trailing edges respectively) as compared to a 90 nm square device in (d),(e) and (f) (with 2 ns, 42 ns and 102 ns trailing edges respectively). Prior to each pulse application, the device is prepared in the predominantly crystalline ($C_A$) state. The standard crystallization pulse used was 2V in amplitude, 1 μs wide with a 150 ns trailing edge. After each programming pulse application, the sample resistance was probed at 50 mV. In fig. 5, for the purpose of clarity in comparing different samples, the color scale is normalized to the maximum measured dc resistance. The maximum resistance was 2 MΩ for the 75nm circular device shown in (a), (b) and (c) and 1.5 MΩ for the 90 nm square device shown in (d), (e) and (f).

The reset operation resistance map in Fig. 5 suggest that the pulse trailing edge (i.e. the cooling rate) plays a key role in determining the critical pulse widths needed to achieve amorphization[13]. Increasing the trailing edge from 2 ns to 42 ns for the circular device (decreasing the cooling rate from ~3 * $10^{11}$ K/s to ~1.5 * $10^{10}$ K/s) results in an increase in the critical pulse width from 10 ns to 60 ns, above which the reset voltage (~0.6V) becomes virtually independent of the applied pulse width. However there exists a narrow range (about 5 ns) below the cited thresholds where the reset operation appears to be possible at the expense of a ~10% increase in the necessary reset pulse amplitude. We note that the failure to reset in the small pulse width range is partially due to the fact that when the pulse widths are less than the trailing edge the pulse shape becomes triangular, attenuating the actual applied pulse amplitude by a factor of pulse width divided by the trailing edge. So in that regime a change in the heating conditions determines the final phase composition rather than the cooling scheme. The set resistance (in the $C_A$ state) was ~1.5 kΩ, suggesting a threshold reset current of ~400 μA. However since the set state resistance is somewhat voltage dependent, from fig. 3 (b) inset this extrapolation for the reset current of the smallest devices is an under-estimate by about 25%. (Comparison of the extrapolated reset currents of the smallest devices with finite-element simulation also suggest that the reset currents are underestimated by ~ 25% as shown in fig. 8).

The corresponding measurements presented in fig. 5 (d), (e) and (f) for the 90 nm square device proved that an intermediate resistance state (a mixed crystalline and amorphous phase) is favorable for a wide range of pulse widths and amplitudes especially for longer pulse trailing edges during the reset operation. We attribute this mixed-phase formation behaviour to the inhomogeneous heating pattern in these devices due to current crowding at the corners, which facilitates the melt-quenching in local spots at the corners (see the discussion on finite-element simulations below). Furthermore, increasing the programming pulse width with a 2 ns trailing edge or increasing the trailing edge for long pulse widths both have the effect of increasing the reset resistance causing a larger fraction of the GST to be melt-quenched whereas the opposite effect is seen in circular devices as shown in in fig. 5 (a), (b) and (c) in agreement with previous reports[19]. In circular devices the observed decrease in reset resistance with increasing pulse width and/or pulse trailing edge is due to the extra time allowed for the atoms to rearrange for the stable-crystalline structure. However in square contacts, this effect seems to be counter-balanced by the inhomogeneous heat flow such that, the atomic rearrangement facilitates the nucleation of a mixed-phase with an intermediate resistance. At larger pulse amplitudes the extra time allowed for heat diffusion from the locally hot spots enable larger volumes of the active region to be melt-quenched, leading to higher reset resistance values. The set resistance for this device was

~700$\Omega$ resulting in a reset current of ~1.7 mA which also evolved with the applied pulse width reaching about 2.3 mA for the 102 ns trailing edge case. Since the contact area of this device is almost twice as that of the 75 nm circle about a factor of 2 increase in the reset currents is expected. However, our observation of a factor or 3 to 4 increase in reset currents in the square device indicates that the threshold reset currents are very sensitive to the exact current distribution as well as the contact area.

Figures 6 (a) shows the corresponding phase diagram for the set operation on the 75 nm circular device with a 2ns trailing edge (similar results were obtained for square devices-not shown). The phase diagram of the set operation suggests that there exists a minimum pulse width of 50 ns above which the crystalline nucleation can be triggered. This characteristic crystallization time for GST is in very good accord with previous findings[3]. The characteristic crystallization time we report here is different from the voltage-dependent set time in the sense that under the application of large pulse amplitudes it is the minimum achievable set-time one can observe in our samples. The set resistance obtained in pulsed switching operation was found to be larger than the resistance of a fresh sample (~1k$\Omega$), which indicated that the pulsed set operation always resulted in a predominantly crystalline state with a residual amorphous content ($C_A$ state). After pulsed switching, a full recovery of the initial resistance (fully crystallized state) was possible in devices with linear dimensions above 100nm by doing a set operation with dc sweeps (sweeping speed ~0.1V/s and without current compliance) suggesting that the pulse induced changes in the set resistance are reversible. This implies that although from Fig. 6 (a) the characteristic crystal nucleation and growth time appears to be of the order of 50 ns, such a time scale is not enough to achieve full crystallization of the GST layer. A study of the set resistance (in the $C_A$ state) as a function of pulse trailing edge for a 1.7V pulse amplitude and a pulse width as large as 0.6 µs (as shown in fig. 6 (b)) shows a clear dependence on the pulse trailing edge. In this case since the applied pulse is expected to cause melting in the contact area of the GST the degree of crystallization along the trailing edge of the pulse can only be enhanced by applying pulses with long trailing edges. Fig.6 (b) suggests that although the crystallization time is about 50 ns, a fully crystalline state can only be formed in a 250 ns time scale. This however, does not present a practical problem for device applications since the relaxation in set resistance is minor compared to the resistance change in between set and reset states.

### FINITE-ELEMENT SIMULATIONS

The results of a 3-D FEM model of the local heating induced by electrical pulses are shown in Fig. 7. This model takes into account the temperature dependence of the thermal conductivities[20,21] and the electrical resistivities[4,21] of both the GST and the WTi layer. The $Al_2O_3$ encapsulation layer has been treated as a perfect thermal insulator. This treatment reduces the simulation time substantially but it also introduces an uncertainty to the quantitative extraction of threshold reset currents, since the thermal conductivity of $Al_2O_3$ and the interfacial thermal conductivity with WTi[22] are not much smaller than that of the GST. Further test-simulations including the $Al_2O_3$ layer showed that, treating the $Al_2O_3$ layer as a perfect thermal insulator introduces a 5% error to the extracted reset currents and does not change the thermal diffusion profile in the active chalcogenide layer qualitatively. The electrical model involves solving the Laplace equation assuming that the electrical conductivity of the GST has no field dependence[4]. The calculated power dissipation acts as the heat source in the thermal model. The calculated temperature of each cell, as determined by the solutions of the heat diffusion equation modifies the electrical resistivity and the thermal conductivity accordingly for the next iteration. The top and bottom electrode surfaces have a fixed thermal boundary condition of T=300K since they act as heat sinks in the cooling process whereas the electrical problem assumes the bottom surface to be grounded with the top electrode surface acting as the electrical pulse source. We note that this simple model does not take into account phase change kinetics (homogeneous-heterogeneous nucleation theories and electrical percolation effects) which are accounted for in a more sophisticated model[4].

Figure 7 (a) and (b) show the temporal evolution of the normalized thermal gradient at the GST-WTi interface for a 100nm circular contact as compared to a 100nm square contact in (d) and (e). During the application of a reset pulse, the local heating pattern clearly shows an inhomogeneous heat distribution in a square contact compared to a circular contact, with preferential local heating at the sharp corners[15]. The inhomogeneous heating scheme in the square contact facilitates the formation of a residual amorphous matrix in the set state leading to more pronounced variations in the snap-back voltage as previously noted in Fig. 3 (a). The resulting temperature distributions in the GST layer at the end of the 50 ns pulse in the circular contact is shown in fig. 7 (c) as compared to the square contact in fig. 7 (f). Since the GST-WTi interface is close to a metal heat sink, the actual hot spot occurs almost half way into the GST layer in both cases with the exception that the hot spot extends all the way to the corners of the contact region in the square device. If the device is quenched to below its glass transition temperature starting from this thermal profile, it results in an amorphous hemisphere surrounded by crystalline GST and, therefore, a slight increase in the device resistance. If the reset voltage level is below the snap-back threshold as is the case for a 75 nm circular device, the highly insulating-amorphous hemisphere allows more concentrated current flow at its surroundings and therefore an accelerated reset switching (amorphization of the GST-WTi interface). Strictly speaking the reset threshold corresponds to the current pulse which causes melt-quenching at the interface with the top electrode (about 300 $^{\circ}$C cooler than the hot spot) rather than a region further into the GST layer[4].

Figure 8 shows the scaling behaviour of reset threshold currents of circular devices (red dots) along with two small square devices we studied with 90 nm and 130 nm sides which had 1.7 mA and 3.8 mA reset currents respectively ( about 50% higher than expected considering the contact areas when compared to circular devices) (simulated value for a 100 nm square device is 1.4 mA). We attribute this slight increase in the reset thresholds of square contacts as compared to the circles to be due to the change in the current distribution and the inhomogeneous heating scheme (also verified by simulations). The resulting behaviour is compared to FEM simulations (blue diamonds) of circular devices . The linear trend in the experimental data ( red solid line) overlaps with the linear trend obtained from simulations (blue solid line) almost over the entire range of contact sizes studied. However we note that deviations from linear scaling for devices with ever smaller dimensions have been observed with TiN electrodes[4] due to a saturation effect attributed to a limitation of having a very small contact area in close proximity with a relative large heat reservoir limiting the ability to get efficient local heating.

**CONCLUSION**

In summary, we have employed a resistance mapping technique to probe the critical phase change conditions of GST phase change memory devices with various contact sizes and circular and rectangular contact geometries. Such measurements reveal that the exact current distribution in the contact region to the active chalcogenides layer plays an essential role in the resulting device performance. Important switching parameters such as the snap-back voltage in consecutive switching events, the reset threshold currents and amorphous state resistance are very sensitive indicators of changes in the current distribution. A non-uniform heating scheme is potentially useful for multiple-bit per cell memory applications since it facilitates the formation of a mixed-phase for a wide range of programming parameters. The scaling behaviour of the reset conditions shows very good agreement with a 3-D FEM model of the electro-thermal physics. Both experimental results and the simulations suggest that the pulsed set and reset operations result in a mixed phase (with a predominantly crystalline or amorphous content) where the amorphous/crystalline volume ratio is highly sensitive to the details of contact geometry and can also be controlled by tuning the cooling rate (pulse trailing edge).

## ACKNOWLEDGEMENTS

The authors would like to thank Sylvia Florez , Vasily Dronov and Neil Smith for useful discussions on high frequency measurements.

## Figure Captions

**Fig 1. (color online)**

**(a)** Schematic drawing of the GST nanopillar device with WTi top and bottom electrodes.

**(b)** Cross-sectional SEM micrograph of a 75nm diameter nanopillar disk.

**Fig 2: (color online)**

IV characteristics of a 600 x 600 nm$^2$ device with blue (bottom) curve corresponding to the amorphous (RESET) state displaying threshold switching (OTS) and memory switching (OMS) to the crystalline state, and the red (upper) curve corresponding to the crystalline (SET) state.

**Fig. 3: (color online)**

Five consecutive IV measurements of a **(a)** 90nm square contact and **(b)** 75 nm circular contact with each measurement starting from the amorphous state (following a pulsed reset operation)..**inset:** the corresponding dc resistance as a function of voltage bias. Dotted lines form an aid to the eye to compare reset state and set state resistance values

**Fig. 4: (color online)**

Pulsed reset operation for a 75 nm circular device with 50 ns pulse width and 2ns trailing edge programming conditions. **inset:** the definition of applied pulse parameters.

**Fig. 5: (color online)**

**(a), (b), (c)** Phase diagrams of the reset operation for a 75 nm circular contact device for 2ns, 22ns and 42 ns trailing edges respectively. The pulse amplitudes are attenuated in the region above the dotted lines due to triangular distortion of the pulse shape.

**(d), (e), (f)** Phase diagrams of the reset operation for a 90 nm square contact device for 2 ns, 42 ns and 102 ns trailing edges respectively. The pulse amplitudes are attenuated in the region above the dotted lines due to triangular distortion of the pulse shape.

**Fig. 6: (color online)**

**(a)** Phase diagram of the set operation for a 75 nm circular contact device for 2ns trailing edge.

**(b)** Set state resistance as a function of pulse trailing edge used in the set operation for a 1.7 V, 0.6 μs wide pulse.

**Fig. 7: (color online)**

Temporal evolution of the simulated temperature gradient profile of a 100nm diameter circular contact **(a)**, **(b)** and a square contact with 100 nm on a side **(d), (e)** (at 40 ns and 50 ns during a 50ns reset pulse respectively).

Cross-sectional view of the simulated spatial temperature distribution in a 100 nm circular contact **(c)** and a 100 nm square contact **(f)** at the end of a 50ns reset pulse.

**Fig. 8: (color online)**

Reset current scaling with contact area of a circular device (red circles experimental data and blue diamonds simulation results with the solid lines being linear fits). (Inset) Non-linear down-scaling of reset currents in the sub-100 nm contact size range.

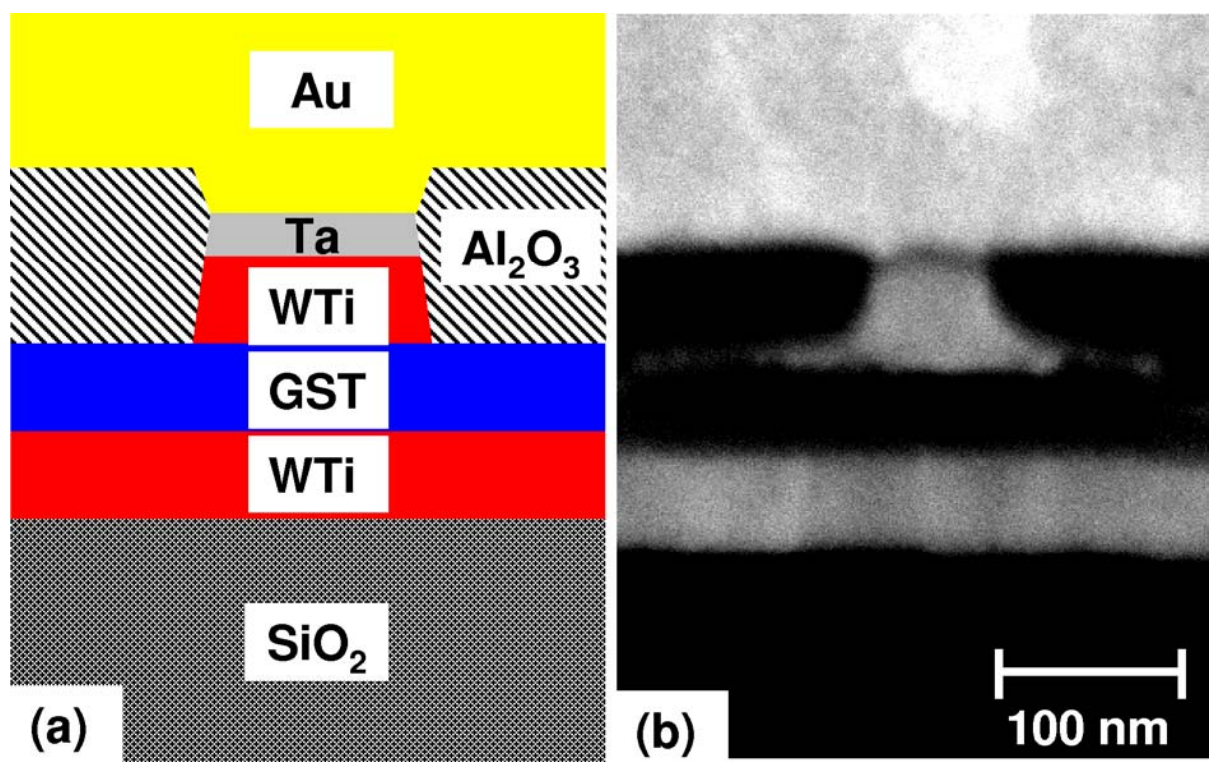

Figure 1
Ozatay et al.

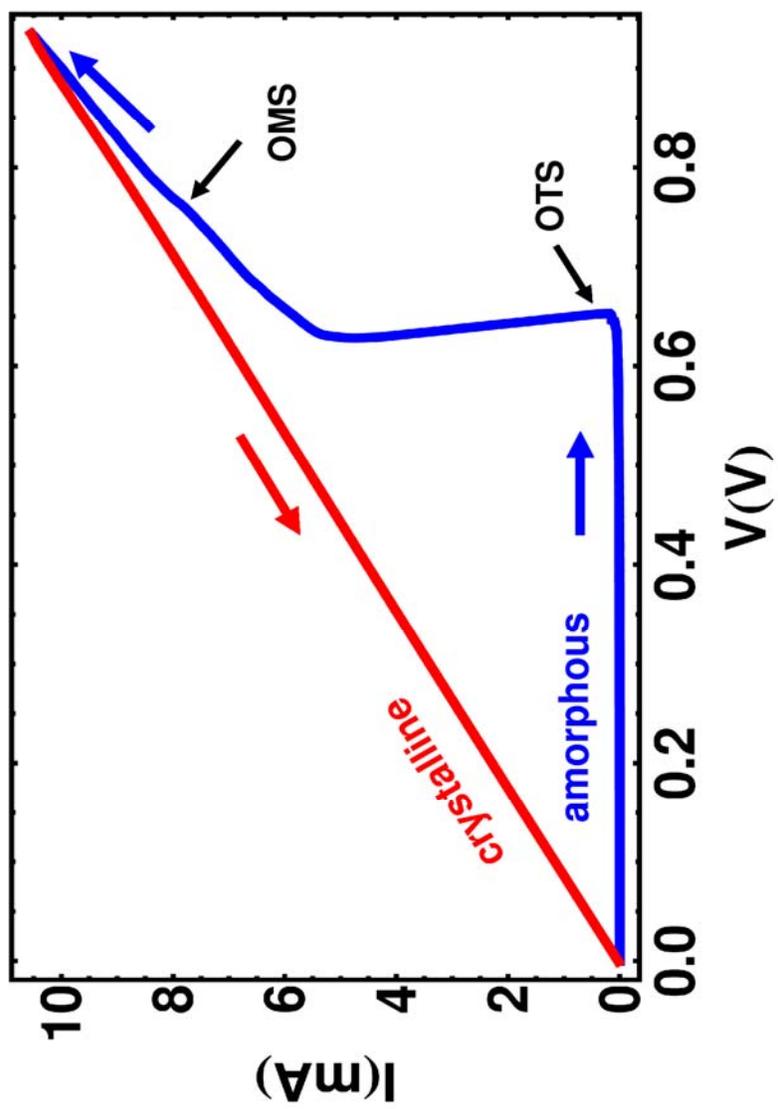

Figure 2
Ozatay et al.

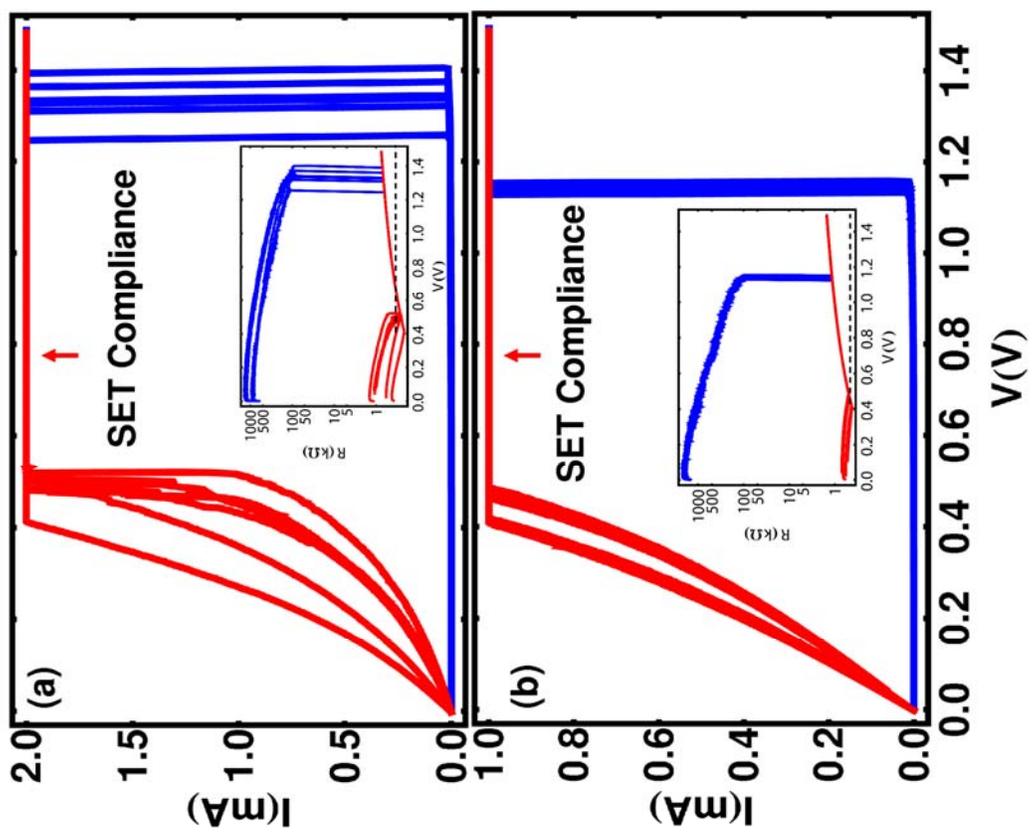

Figure 3
Ozatay et al.

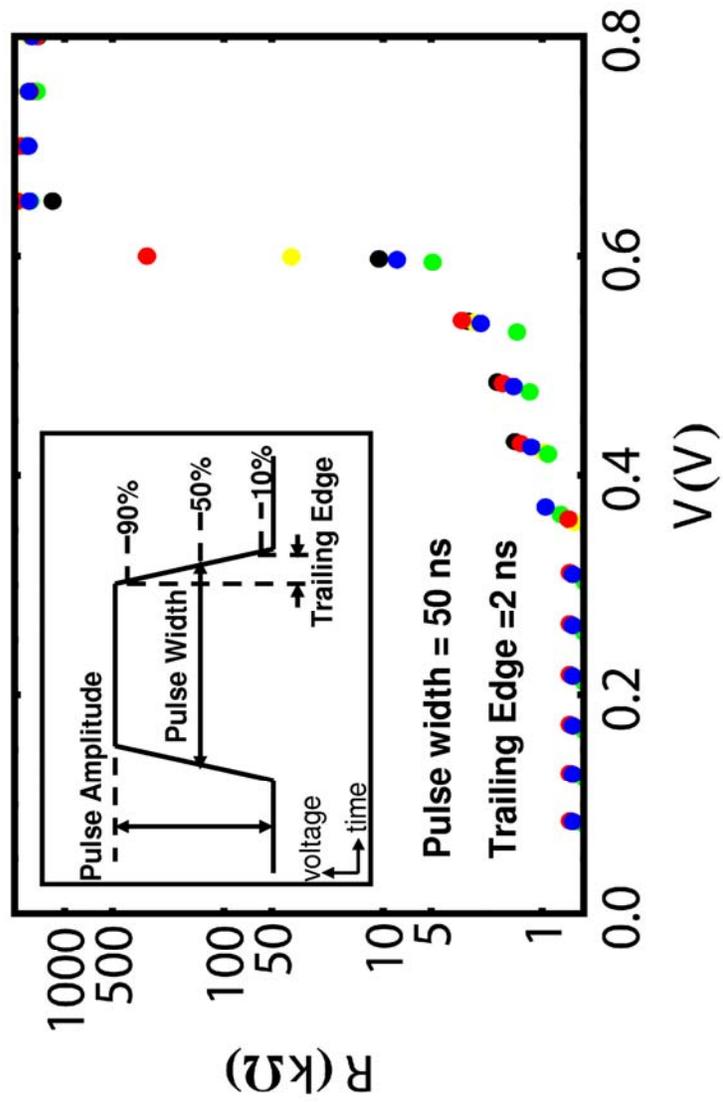

Figure 4
Ozatay et al.

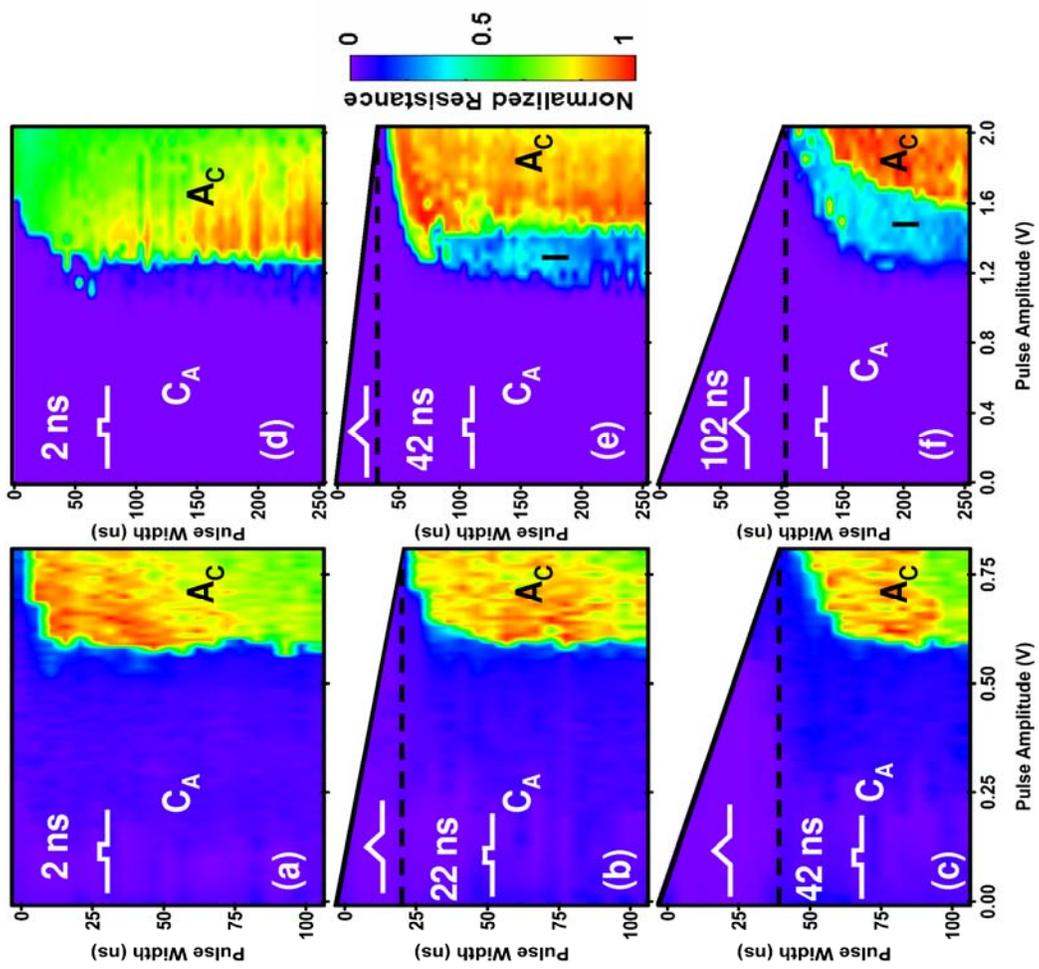

Figure 5
Ozatay et al.

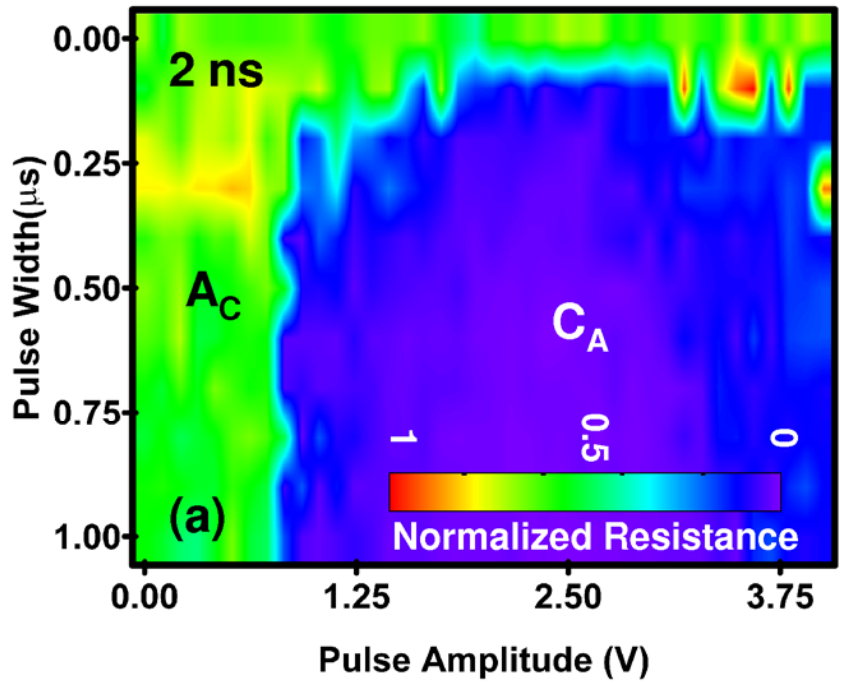

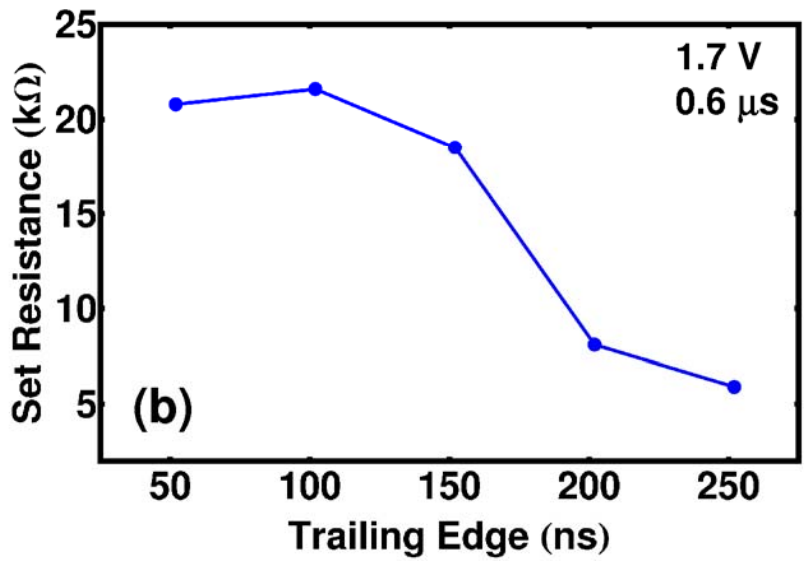

Figure 6
Ozatay et al.

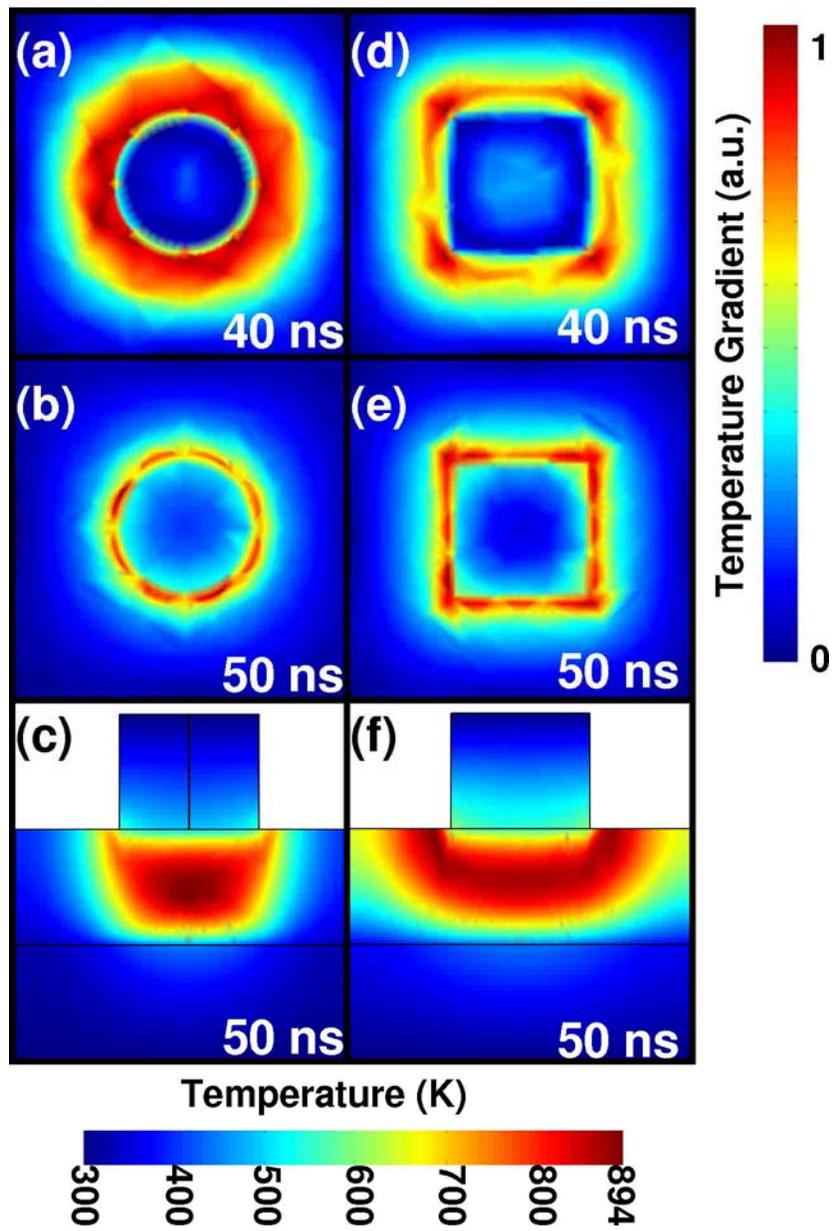

Figure 7
Ozatay et al.

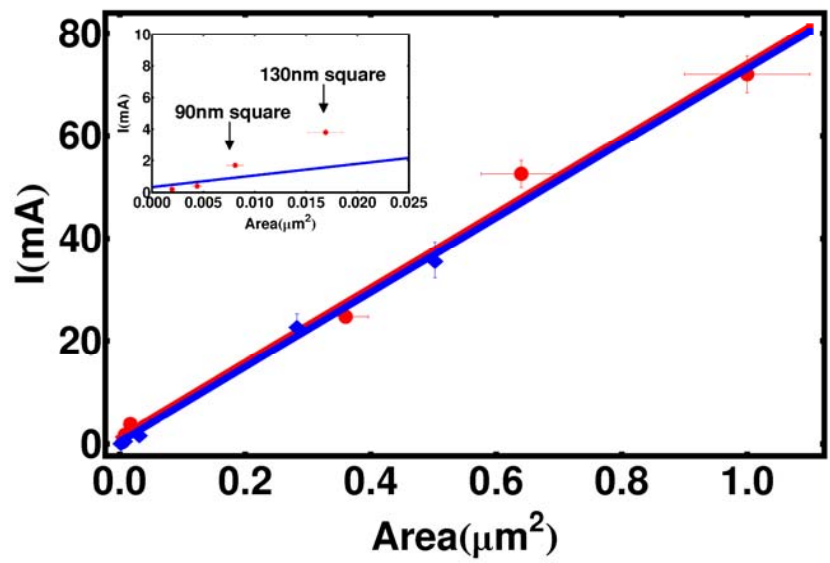

Figure 8
Ozatay et al.